\documentclass[cameraready]{Interspeech}

\title{How~Auditory~Knowledge~in~LLM~Backbones~Shapes~Audio~Language~Models: A Holistic Evaluation}

\author[affiliation={1}]{Ke-Han}{Lu}
\author[affiliation={2}]{Szu-Wei}{Fu}
\author[affiliation={2}]{Chao-Han Huck}{Yang}
\author[affiliation={2}]{Zhehuai}{Chen}
\author[affiliation={2}]{Sung-Feng}{Huang}
\author[affiliation={1}]{Chih-Kai}{Yang}
\author[affiliation={1}]{Yi-Cheng}{Lin}
\author[affiliation={1}]{Chi-Yuan}{Hsiao}
\author[affiliation={1}]{Wenze}{Ren}
\author[affiliation={1}]{En-Pei}{Hu}
\author[affiliation={1}]{Yu-Han}{Huang}
\author[affiliation={1}]{An-Yu}{Cheng}
\author[affiliation={1}]{Cheng-Han}{Chiang}
\author[affiliation={3}]{Yu}{Tsao}
\author[affiliation={2}]{Yu-Chiang Frank}{Wang}
\author[affiliation={1}]{Hung-yi}{Lee}

\address{
    $^1$ National Taiwan University, Taiwan \quad $^2$ NVIDIA \quad
    $^3$ Academia Sinica, Taiwan
}

\email{d12942024@ntu.edu.tw, hungyilee@ntu.edu.tw}


\usepackage{comment}
\usepackage{multirow}
\usepackage{booktabs}
\usepackage{graphicx}
\usepackage{amsmath} 
\usepackage{colortbl}
\usepackage{xcolor}
\usepackage{cite}
\usepackage{subcaption}

\begin{document}

\maketitle

\begin{abstract}

Large language models (LLMs) have been widely used as knowledge backbones of Large Audio Language Models (LALMs), yet how much auditory knowledge they encode through text-only pre-training and how this affects downstream performance remains unclear. We study this gap by comparing different LLMs under two text-only and one audio-grounded setting: (1) direct probing on AKB-2000, a curated benchmark testing the breadth and depth of auditory knowledge; (2) cascade evaluation, where LLMs reason over text descriptions from an audio captioner; and (3) audio-grounded evaluation, where each LLM is fine-tuned into a Large Audio Language Model (LALM) with an audio encoder. Our findings reveal that auditory knowledge varies substantially across families, and text-only results are strongly correlated with audio performance. Our work provides empirical grounding for a comprehensive understanding of LLMs in audio research.\footnote{https://kehanlu.github.io/AKB}

\end{abstract}

\section{Introduction}
Large Language Models (LLMs) trained on massive text corpora have demonstrated a remarkable ability to internalize world knowledge across diverse domains, from general reasoning to specialized technical fields~\cite{dubey2024llama,yang2024qwen25, yang2025qwen3, comanici2025gemini,hurst2024gpt,singh2025openai,anthropic2025sonnet45, abdin2024phi, olmo2025olmo}.
Among the various types of knowledge, the linguistic representation of auditory experiences is of particular interest. Humans routinely describe auditory perception through text: we write that a violin sounds warm, that a siren grows louder as it approaches, or that a speaker's tone conveys anger. These textual descriptions allow a reader to reason about sounds without hearing them. It is therefore natural to hypothesize that LLMs have acquired substantial auditory knowledge through text-only training alone.

In the current research landscape, LLMs predominantly empower audio understanding systems through several paradigms. First, an LLM serves as the cognitive and knowledge backbone of a Large Audio Language Model (LALM), paired with an audio encoder and jointly fine-tuned on audio-oriented data to bridge acoustic features into its pre-existing linguistic space~\cite{chu2024qwen2, xu2025qwen3, gong2023joint, tang2024salmonn,lu2025desta25, lu24c_interspeech,desta2,hu2024wavllm,ghosh-etal-2024-gama,pmlr-v267-ghosh25b,goel2025audio, abouelenin2025phi}. Alternatively, an LLM can operate within a cascade pipeline, where a specialized audio-to-text module first converts the input into text, which the LLM subsequently interprets to generate a response~\cite{ma2025omni,rong2025audiogenie,taheri2025sarlm,kuan2024speech}. Second, the LLM often acts as a synthetic data engine to curate audio-centric training sets, for example by rephrasing audio descriptions~\cite{mei2023wavcaps,lu24c_interspeech,ma2025omni} or synthesizing audio instruction-tuning datasets~\cite{desta2,lu2025desta25,gong2023joint,hu2024wavllm,xie2025audio,goel2025audio}. Crucially, in these roles, the depth and accuracy of the auditory knowledge encoded within the text-only LLM serve as a fundamental determinant of the resulting system's performance.

\begin{figure*}[t]
    \centering
    \includegraphics[width=0.99\linewidth]{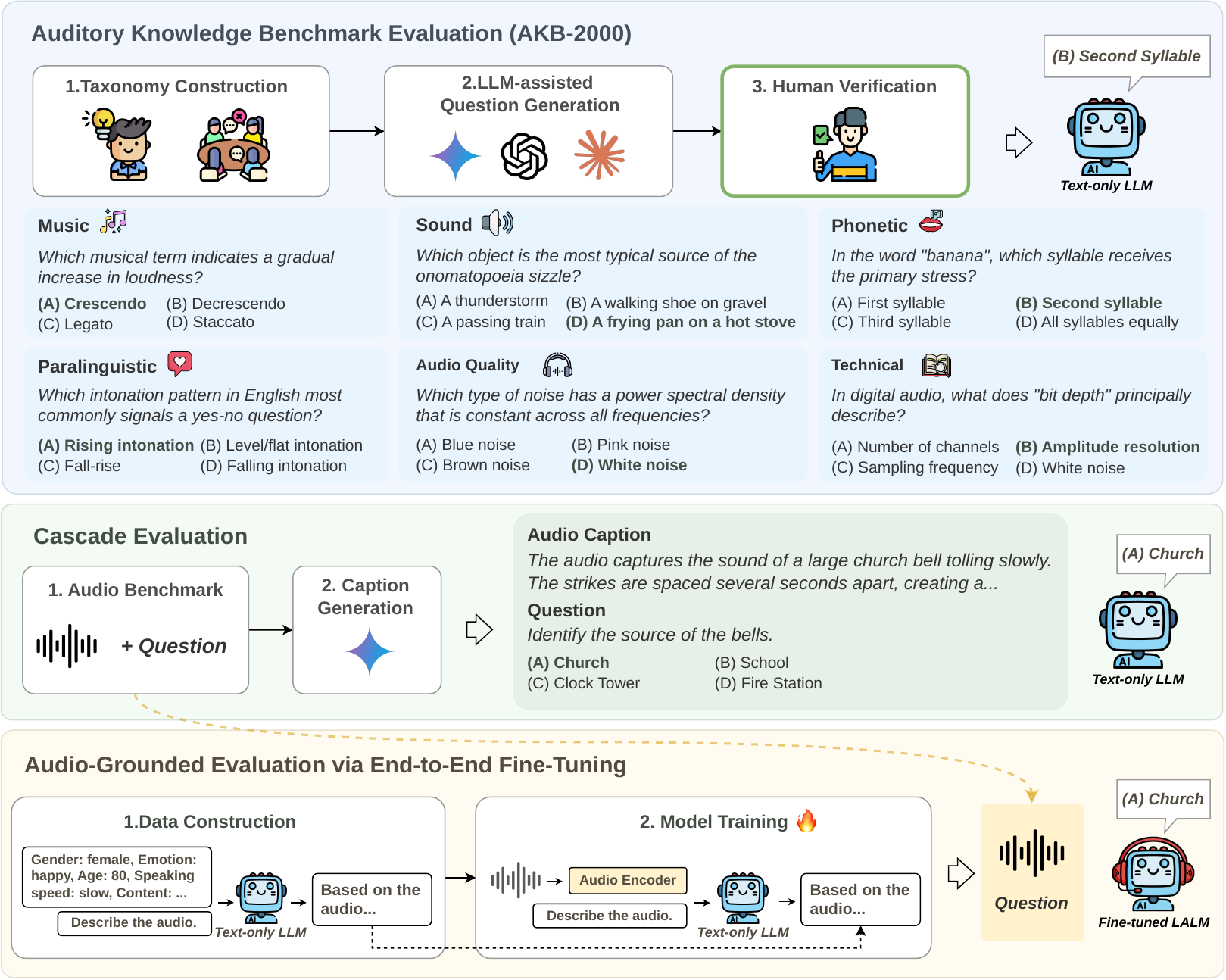}
    \caption{Overview of the three evaluations introduced in this work. \textbf{(Top) AKB-2000 construction pipeline:} a two-level taxonomy guides LLM-assisted question generation, followed by human verification. \textbf{(Middle) Cascade evaluation:} a captioner converts audio to text descriptions fed to a text-only LLM. \textbf{(Bottom) Audio-grounded evaluation:} each LLM is fine-tuned into a LALM using the DeSTA self-distillation framework and evaluated with audio input.}
    \label{fig:figure1}
\end{figure*}

However, most existing LALM studies select a single LLM, devoting their analysis to architectural design, training strategy, or audio encoder choice, leaving the role of the LLM backbone unclear.
For example, Llama~\cite{lu24c_interspeech,desta2,lu2025desta25,ghosh-etal-2024-gama,hu2024wavllm,yang2024building} and Qwen~\cite{chu2023qwen,chu2024qwen2,pmlr-v267-ghosh25b,xu2025qwen3} are the two most frequently adopted LLM backbones in existing LALMs, yet the choice of backbone is rarely justified or evaluated on the basis of the LLM's own auditory knowledge.
We argue that LLMs trained on distinct corpora with varying training recipes likely manifest markedly different levels of auditory understanding, and that a model with a richer internal representation of sound may hold an inherent advantage in multimodal adaptation. Consequently, it remains unclear how much auditory knowledge current LLMs actually possess and to what extent this knowledge influences their multimodal adaptation.

In this work, we present a systematic evaluation to investigate the auditory knowledge encoded in text-only LLMs and their relative strengths. As illustrated in Figure~\ref{fig:figure1}, we introduce two text-only and one multimodal evaluation. In the text-only settings, we assess auditory knowledge with two paradigms. The first is \textbf{direct auditory knowledge evaluation}, where we evaluate different LLMs on AKB-2000, an auditory question-answering benchmark we have curated that covers a wide range of topics in audio research, spanning 6 categories including Music, Sound, Paralinguistic, Phonetic, Audio Quality and Technical knowledge. The second is \textbf{cascade evaluation}, where an audio captioner translates audio samples from existing audio benchmarks into detailed descriptions for the LLM to answer the original question.
%
The third is \textbf{audio-grounded evaluation}, where we fine-tune each LLM into an end-to-end LALM by pairing it with an audio encoder, following the self-distillation framework from DeSTA~\cite{desta2,lu2025desta25}. 
This setup provides a controlled environment to directly assess whether inherent auditory knowledge in text-only LLMs transfers to better audio understanding after multimodal adaptation.

We evaluate 12 open-weight LLMs spanning 4 model families (Qwen~\cite{yang2024qwen25,yang2025qwen3}, Llama~\cite{touvron2023llama,dubey2024llama}, OLMo~\cite{olmo2025olmo}, Phi~\cite{abdin2024phi,abouelenin2025phi}) across different model generations, training stages, and parameter scales. We also include 5 proprietary models such as GPT~\cite{hurst2024gpt,singh2025openai}, Gemini~\cite{comanici2025gemini}, and Claude~\cite{anthropic2025sonnet45} as strong baselines.
Our comprehensive evaluation reveals several key findings. First, auditory knowledge varies substantially across model families, with Qwen consistently outperforming Llama in most evaluated settings. When both models are fine-tuned with an identical training recipe, the choice of the base LLM alone can result in over a 10\% absolute performance difference in the resulting LALM. Second, there is a strong positive correlation between text-only evaluation and audio-grounded evaluation. This indicates that text-only benchmarks can serve as a reliable and lightweight proxy for selecting backbone models prior to expensive multimodal training. Furthermore, we identify that LLMs consistently struggle with phonological tasks, highlighting the inherent limitations of text-only pre-training. Finally, we observe that a simple cascade pipeline using captioned text can match or even surpass several state-of-the-art end-to-end LALMs, suggesting that current end-to-end systems are bottlenecked by the audio encoder, leaving the LLM's inherent auditory reasoning capability underutilized.


Our contributions can be summarized as follows:
\begin{itemize}
    \item We provide a holistic evaluation of 12 open-weight LLMs through the lens of audio understanding systems, providing actionable takeaways that can help select the optimal LLM for fine-tuning an LALM.
    \item We introduce AKB-2000, a curated auditory knowledge benchmark with 2,000 questions covering 6 categories and 48 subcategories in audio research.
    \item We will release the code, benchmarks, and model checkpoints to ensure transparency and to support future research.
\end{itemize}

\section{Related Work}
\subsection{Audio Understanding Systems}

LLMs have become foundational in audio research, underpinning significant advancements in automatic speech recognition~\cite{chen2023hyporadise,10389705}, text-to-speech~\cite{wang2023neural,du2024cosyvoice}, and spoken dialogue systems~\cite{defossez2024moshi,fang2025llamaomni,rubenstein2023audiopalm,arora2025landscapespokenlanguagemodels,yang2024building,hsiao25_interspeech}. In this work, we focus on audio understanding systems, which aim to bridge raw acoustic signals with linguistic reasoning to execute diverse, open-ended tasks, necessitating both robust perception of complex acoustic scenes and the semantic capacity to interpret nuanced auditory cues.

These systems can be broadly categorized into two paradigms, namely end-to-end LALMs and modular agentic systems. End-to-end LALMs couple an audio encoder with an LLM backbone via a modality connector, with representative models including LTU~\cite{gong2023joint}, SALMONN~\cite{tang2024salmonn}, Qwen-Audio~\cite{chu2023qwen,chu2024qwen2,xu2025qwen3}, Phi-4-mm~\cite{abouelenin2025phi}, DeSTA~\cite{lu24c_interspeech,desta2,lu2025desta25}, and Audio Flamingo~\cite{pmlr-v235-kong24a,pmlr-v267-ghosh25b,goel2025audio}. By mapping acoustic features directly into the LLM's latent space through multimodal instruction tuning, these models leverage the LLM's internal knowledge to support flexible multimodal interaction. Beyond architectural design, the development of these systems increasingly relies on LLMs for data curation, ranging from synthesizing open-ended question-answer pairs to augmenting audio captions for pre-training~\cite{desta2,lu2025desta25,gong2023joint,hu2024wavllm,xie2025audio,goel2025audio}. An emerging trend in this direction further incorporates self-distillation into the data construction process~\cite{fathullah2023audiochatllama,wang2023blsp,desta2,lu2025desta25,fujita25b_interspeech, xie2025enhancing}, emphasizing the LLM's inherent auditory reasoning capacity to enable zero-shot generalization to unseen tasks without task-specific fine-tuning, as demonstrated by frameworks such as DeSTA~\cite{desta2,lu2025desta25}.

Modular agentic systems~\cite{ma2025omni,rong2025audiogenie,taheri2025sarlm,kuan2024speech}, by contrast, employ a cascade pipeline in which a specialized audio-to-text module such as an ASR system or audio captioner first converts the input signal into an intermediate textual representation, which an LLM subsequently interprets to generate a response. While this approach offers greater interpretability and avoids the cost of multimodal training, its performance is inherently bounded by the descriptive granularity of the intermediate text. End-to-end LALMs, on the other hand, face persistent challenges in cross-modal alignment and catastrophic forgetting during fine-tuning~\cite{lu2025speechifeval}.

Despite their architectural differences, both paradigms share a common assumption that the underlying LLM possesses sufficient auditory knowledge to support downstream reasoning. How much such knowledge is actually encoded through text-only pre-training, and how it translates to multimodal performance, remains an open empirical question that directly motivates our work.

\subsection{Evaluating Auditory Knowledge and Capabilities}
The evaluation of audio understanding systems has evolved from task-specific benchmarks~\cite{panayotov2015librispeech,gemmeke2017audio,piczak2015dataset,yang21c_interspeech} toward holistic, instruction-following assessments~\cite{huang2025dynamicsuperb,sakshi2025mmau,ma2025mmar,yang2025sakuramultihopreasoninglarge,yang-etal-2025-towards-holistic,yang2025audiolens,wang2024audiobench,lu2025speechifeval}. For instance, MMAU~\cite{sakshi2025mmau} assesses multitask understanding across sound, music, and speech, and MMAR~\cite{ma2025mmar} further requires deeper reasoning beyond surface-level perception. Although these benchmarks have been widely adopted for system-level comparison, they conflate multiple factors simultaneously: audio encoding quality, training data coverage, and the LLM's internal knowledge. As a result, when a performance gap is observed, it is difficult to determine whether the cause is a weak audio encoder, insufficient training data, or a fundamental deficiency in the LLM's auditory knowledge.

A complementary line of research has begun to probe whether LLMs acquire auditory knowledge implicitly through text pretraining. Prior work has approached this via representation probing~\cite{ngo-kim-2024-language}, retrieval- and generation-based auditory knowledge augmentation~\cite{ok2025audiobert, yoo2025imagine}, and direct question-answering on low-level acoustic attributes such as pitch, loudness, and animal sound recognition~\cite{ok2025audiobert, ok2025auditorybench++}. However, these studies are limited to basic sound events and coarse acoustic properties, leaving open the question of whether LLMs possess the broader auditory knowledge required for general-purpose audio understanding.

Our work addresses this gap along three dimensions. First, we systematically probe LLMs across a broader and more diverse set of auditory tasks and domains than previously examined, establishing AKB-2000 as a new benchmark for evaluating auditory knowledge in text-only settings. Second, we extend this evaluation to a cascade setting, testing whether LLMs can apply their encoded auditory knowledge to reason over real audio questions represented as text, and examining how this capability varies across model families. Third, we analyze how both forms of text-only knowledge correlate with performance after audio fine-tuning, offering the first direct empirical link between an LLM's text-based auditory knowledge and its audio-grounded understanding capability.

\section{Method}

We introduce three complementary evaluations that investigate the auditory knowledge encoded in different LLMs across two text-only and one multimodal setting. In the text-only settings, we evaluate LLMs on two paradigms. The first is direct question answering on audio-related common sense and factual knowledge (Section~\ref{sec:aqa}). The second is cascade evaluation, where LLMs answer questions from existing audio benchmarks given textual descriptions produced by a strong captioner (Section~\ref{sec:captioner}). In the multimodal settings, we fine-tune each LLM into a general-purpose LALM and evaluate with actual audio inputs from the same audio benchmarks (Section~\ref{sec:audio-finetuning}). Across all three evaluations, we isolate the LLM backbone as the sole variable, so that observed performance differences can be attributed to the auditory knowledge each LLM encodes. A model that consistently falls short across all three settings may lack sufficient auditory knowledge to serve as a robust foundation for downstream audio systems.

\subsection{Text-only Auditory Knowledge Benchmark Evaluation}
\label{sec:aqa}

To evaluate whether LLMs possess specific auditory concepts, we curate the Auditory Knowledge Benchmark (AKB-2000), a 2,000-question multiple-choice benchmark designed to directly test the breadth and depth of factual knowledge and common sense required for a general-purpose audio system.

Figure~\ref{fig:figure1}-Top illustrates the data collection process and representative examples from each category. We first manually construct a two-level taxonomy consisting of 6 top-level categories and 48 fine-grained subcategories, namely Sound, Paralinguistic, Phonetic, Music, Audio Quality, and Technical Knowledge. This taxonomy spans the major domains of audio research and provides a comprehensive evaluation scope. We primarily focus on auditory concepts that go beyond pure content understanding, since content-level tasks such as general question answering can already be evaluated with existing text-only benchmarks~\cite{hendrycks2021measuring,wang2024mmlupro,srivastava2023beyond}.

Based on the taxonomy, we write detailed topic-specific guidelines for each subcategory, then generate four-option multiple-choice questions with the assistance of three proprietary LLMs (GPT-5, Gemini-2.5-Pro, and Claude-Sonnet-4.5), each producing multiple candidate questions that follow the taxonomy and question design guidelines. Each candidate question is independently verified by two human annotators with audio background who assess correctness, clarity, and the plausibility of distractor options. Only questions where both annotators agree are retained. The final benchmark contains 2,000 verified questions approximately uniformly distributed across all 48 subcategories, ensuring balanced coverage of the taxonomy.

As shown in Figure~\ref{fig:figure1}, our questions range from perceptual knowledge acquired through daily experience, such as associating onomatopoeia with their sound sources and recognizing stress patterns in words, to technical concepts that require domain expertise, such as understanding properties of different noise types and music theory. This breadth allows us to profile the auditory knowledge landscape of each LLM.

\subsection{Text-only Cascade Evaluation}
\label{sec:captioner}

Beyond direct knowledge probing through question answering, which measures what general auditory knowledge an LLM has encoded, we further evaluate LLMs in a cascade pipeline to test whether they can apply this knowledge to interpret and reason about real audio questions.

We adopt MMAU~\cite{sakshi2025mmau} and MMAR~\cite{ma2025mmar} as our evaluation benchmarks, which together cover both recognition and reasoning capabilities expected of a general-purpose audio understanding system. While both benchmarks provide cascade baselines pairing audio captioners with proprietary LLMs, they treat this setting as a naive baseline for end-to-end LALMs rather than systematically comparing across LLMs. We extend this setup to a broader set of LLMs and also vary the captioner to examine how caption quality interacts with LLM capability. 

As depicted in Figure~\ref{fig:figure1}-Middle, given the audio and questions from the audio benchmark, we first prompt Gemini-2.5-Pro (Audio) to produce a detailed textual description for each audio sample that captures salient acoustic properties, sound sources, temporal structure, spoken content, and speaking style. Then, each LLM is asked to answer the audio-related question based on the textual information.

These two text-only evaluations serve complementary roles. AKB-2000 tests auditory knowledge through human-curated questions spanning a broad taxonomy, including factual and technical knowledge that is difficult to assess through audio examples alone. Cascade evaluation, in contrast, tests whether LLMs can apply this knowledge to reason over real audio questions.

\subsection{Audio-Grounded Evaluation via End-to-End Fine-Tuning}
\label{sec:audio-finetuning}

The text-only evaluations above reveal what LLMs know about audio through text alone, but leave open whether this knowledge translates to better performance when real audio waveforms replace text as input. To answer this question, we fine-tune each LLM into an LALM by pairing it with an audio encoder and jointly fine-tuning on audio instruction-tuning data. By comparing different LLMs, we can investigate whether the auditory knowledge identified in the text-only settings transfers to an audio-grounded evaluation, and whether a stronger text-only LLM yields a stronger LALM when processing real audio waveforms. We evaluate the resulting LALMs on MMAU and MMAR, the same benchmarks used in the cascade evaluation, using actual audio waveforms as input. 

%
To fine-tune an LLM into an LALM, we adopt the self-distillation framework from DeSTA~\cite{lu2025desta25}, which consists of two stages as shown in Figure~\ref{fig:figure1}-Bottom. In the first stage, the LLM reads textual metadata associated with each audio sample, such as attribute labels or audio descriptions, and generates a response to a randomly sampled prompt (e.g., ``Describe the audio.''). In the second stage, the raw audio waveform replaces the textual metadata as input. The audio is processed by an audio encoder and projected into the LLM input space through a modality connector, and the model is optimized end-to-end to reproduce the response generated in the first stage.

This framework is particularly suited to our study because the backbone LLM shapes the resulting LALM through two distinct pathways. On the data side, each LLM generates its own training targets from textual audio descriptions, so an LLM with richer auditory knowledge produces more accurate and informative supervision signals. On the model side, since the training targets are generated by the backbone LLM itself, the optimization objective is inherently closed with the model's existing knowledge and generation style, which has been shown to preserve the original capabilities of the backbone during continued training~\cite{desta2,lu2025desta25,wang2023blsp,fathullah2023audiochatllama}.


\section{Experimental Setup}
\subsection{Evaluated LLMs}

We select 12 open-weight instruction-tuned LLMs, covering four model families: Qwen~\cite{yang2024qwen25,yang2025qwen3}, Llama~\cite{touvron2023llama,dubey2024llama}, Phi~\cite{abdin2024phi, abouelenin2025phi}, and OLMo~\cite{olmo2025olmo}. The selection spans parameter scales from 4B to 14B.

Qwen and Llama are among the most frequently used LLM backbones in existing audio research. Qwen serves as the backbone for Qwen-Audio~\cite{chu2023qwen,chu2024qwen2,xu2025qwen3} and AudioFlamingo~\cite{pmlr-v267-ghosh25b,goel2025audio}, while Llama underpins systems such as DeSTA~\cite{lu24c_interspeech,desta2,lu2025desta25}, GAMA~\cite{ghosh-etal-2024-gama}, and WavLLM~\cite{hu2024wavllm}. We include multiple generations within these families, specifically Llama-2-7B, Llama-3-8B, and Llama-3.1-8B from the Llama family, and Qwen2.5-7B, Qwen3-4B, Qwen3-8B, and Qwen3-14B from the Qwen family, to examine how auditory knowledge evolves across model generations. Phi-4-14B and Phi-4-mini-4B are included as the Phi family also has a multimodal audio variant, Phi-4-mm~\cite{abouelenin2025phi}. 
For OLMo-3, we include three checkpoints from the same training pipeline, namely OLMo-3-7B-SFT, OLMo-3-7B-DPO, and OLMo-3-7B (Instruct). As the only fully open-source model family with transparent training data and procedure, OLMo serves as a valuable open-source reference point in our evaluation. In addition to open-weight models, we evaluate five proprietary LLMs as reference points: GPT-5, GPT-4o, Gemini-2.5-Pro, Gemini-2.0-Flash, and Claude-Sonnet-4.5.

\subsection{Fine-Tuning Configuration}

In the audio-grounded evaluation, we fine-tune 8 open-weight LLMs: Qwen3-14B, Qwen3-8B, Qwen3-4B, Qwen2.5-7B, Llama-3.1-8B, Phi-4-14B, Phi-4-mini-4B, and OLMo-3-7B. These models are selected to cover diverse model families at each parameter scale, enabling cross-family comparison at matched sizes.

Each LLM is fine-tuned into an LALM following the DeSTA framework~\cite{desta2,lu2025desta25}, using an identical training recipe. We use DeSTA-AQA500K as the source training data, which is a collective annotation from publicly available datasets containing 404 hours of speech, 329 hours of sound events, and 144 hours of music. Each sample is associated with an audio file, a text field providing textual metadata of the audio content (i.e., seed description), and a text prompt. Following the self-distillation procedure described in Section~\ref{sec:audio-finetuning}, we feed the seed description and prompt to each LLM to generate model-specific training targets. All LLMs share the same source data, and only the generated responses differ.

For model architecture, we use Whisper-large-v3~\cite{radford2023robust} as the audio encoder and a 6-layer Q-Former~\cite{pmlr-v202-li23q, desta2, lu2025desta25} as the modality connector. We freeze both the audio encoder and the LLM parameters throughout training, leaving the modality connector as the only trainable component. Under this setup, the connector learns to project audio representations into a form that the frozen LLM can interpret, providing a stricter test of pre-existing auditory knowledge than full fine-tuning, where the LLM could potentially compensate for knowledge gaps through parameter updates.

All models are trained for 10 epochs with a learning rate of 1e-4 and 2,000 warm-up steps on 2 NVIDIA H100 GPUs, with a per-device batch size of 12 and gradient accumulation of 4, yielding a global batch size of 96. For Qwen3 models, we disable the thinking mode during both data generation and inference to ensure consistent comparison with non-reasoning models. Training is conducted using the official DeSTA codebase.


\begin{table*}[t]
\centering
\caption{
Text-only Auditory Knowledge Benchmark (AKB-2000) evaluation accuracy (\%) across six categories. \colorbox{cyan!20!white}{Shading} indicates relative ranking among open-weight models (darker $=$ higher); values in parentheses show the gap from Gemini-2.5-Pro.
}
\label{tab:aqa_results}
\begin{tabular}{l|c|cccccc}
\toprule
\textbf{Model} & \textbf{Avg.} & Sound & Paralinguistic & Phonetic & Music & Quality  & Technical  \\
\midrule
\multicolumn{8}{l}{\textit{Proprietary LLMs}} \\
Gemini-2.5-Pro & 96.05 & 96.37 & 96.46 & 94.93 & 95.76 & 97.45 & 95.35 \\
GPT-5 & 94.35 & 95.16 & 94.44 & 93.55 & 95.51 & 94.90 & 92.44 \\
Claude-Sonnet-4.5 & 95.70 & 95.16 & 96.97 & 91.71 & 96.01 & 95.41 & 96.22 \\
Gemini-2.0-Flash & 91.85 & 93.15 & 92.42 & 88.94 & 92.77 & 94.39 & 89.24 \\
GPT-4o & 92.90 & 95.97 & 93.94 & 90.32 & 95.01 & 93.88 & 87.50 \\
\midrule
\multicolumn{8}{l}{\textit{Open-weight LLMs}} \\
Qwen3-14B & \cellcolor{cyan!46!white}85.05\,{\scriptsize(-11.00)} & \cellcolor{cyan!50!white}90.73 & \cellcolor{cyan!46!white}85.35 & \cellcolor{cyan!46!white}76.96 & \cellcolor{cyan!46!white}88.78 & \cellcolor{cyan!50!white}88.27 & \cellcolor{cyan!46!white}79.36  \\
Qwen3-8B & \cellcolor{cyan!34!white}78.95\,{\scriptsize(-17.10)} & \cellcolor{cyan!34!white}82.66 & \cellcolor{cyan!34!white}79.80 & \cellcolor{cyan!34!white}68.20 & \cellcolor{cyan!34!white}85.04 & \cellcolor{cyan!38!white}82.65 & \cellcolor{cyan!34!white}72.38  \\
Qwen3-4B & \cellcolor{cyan!42!white}82.00\,{\scriptsize(-14.05)} & \cellcolor{cyan!38!white}85.48 & \cellcolor{cyan!46!white}85.35 & \cellcolor{cyan!42!white}69.59 & \cellcolor{cyan!38!white}86.78 & \cellcolor{cyan!46!white}85.71 & \cellcolor{cyan!42!white}73.84  \\
Qwen2.5-7B & \cellcolor{cyan!38!white}80.70\,{\scriptsize(-15.35)} & \cellcolor{cyan!42!white}87.10 & \cellcolor{cyan!38!white}81.99 & \cellcolor{cyan!38!white}69.12 & \cellcolor{cyan!42!white}87.28 & \cellcolor{cyan!34!white}81.63 & \cellcolor{cyan!38!white}72.97  \\
\midrule
Llama-3.1-8B & \cellcolor{cyan!18!white}68.10\,{\scriptsize(-27.95)} & \cellcolor{cyan!22!white}73.39 & \cellcolor{cyan!14!white}69.53 & \cellcolor{cyan!30!white}58.53 & \cellcolor{cyan!22!white}75.56 & \cellcolor{cyan!22!white}72.96 & \cellcolor{cyan!14!white}56.40  \\
Llama-3-8B & \cellcolor{cyan!30!white}73.45\,{\scriptsize(-22.60)} & \cellcolor{cyan!30!white}79.44 & \cellcolor{cyan!30!white}74.58 & \cellcolor{cyan!18!white}55.76 & \cellcolor{cyan!30!white}81.55 & \cellcolor{cyan!34!white}81.63 & \cellcolor{cyan!26!white}64.24  \\
Llama-2-7B & \cellcolor{cyan!5!white}45.90\,{\scriptsize(-50.15)} & \cellcolor{cyan!5!white}51.61 & \cellcolor{cyan!5!white}44.61 & \cellcolor{cyan!5!white}39.17 & \cellcolor{cyan!5!white}52.62 & \cellcolor{cyan!5!white}44.90 & \cellcolor{cyan!5!white}40.99  \\
\midrule
Phi-4-14B & \cellcolor{cyan!50!white}86.35\,{\scriptsize(-9.70)} & \cellcolor{cyan!46!white}89.92 & \cellcolor{cyan!50!white}88.22 & \cellcolor{cyan!50!white}79.26 & \cellcolor{cyan!50!white}89.53 & \cellcolor{cyan!46!white}85.71 & \cellcolor{cyan!50!white}81.69  \\
Phi-4-mini-4B & \cellcolor{cyan!26!white}70.00\,{\scriptsize(-26.05)} & \cellcolor{cyan!26!white}75.00 & \cellcolor{cyan!22!white}71.04 & \cellcolor{cyan!22!white}56.22 & \cellcolor{cyan!22!white}75.56 & \cellcolor{cyan!18!white}71.94 & \cellcolor{cyan!30!white}65.70  \\
\midrule
OLMo-3-7B & \cellcolor{cyan!22!white}69.05\,{\scriptsize(-27.00)} & \cellcolor{cyan!22!white}73.39 & \cellcolor{cyan!26!white}71.21 & \cellcolor{cyan!26!white}57.60 & \cellcolor{cyan!22!white}75.56 & \cellcolor{cyan!26!white}75.51 & \cellcolor{cyan!22!white}58.14 \\
OLMo-3-7B-DPO & \cellcolor{cyan!14!white}67.30\,{\scriptsize(-28.75)} & \cellcolor{cyan!14!white}71.37 & \cellcolor{cyan!18!white}70.03 & \cellcolor{cyan!18!white}55.76 & \cellcolor{cyan!26!white}75.81 & \cellcolor{cyan!14!white}71.43 & \cellcolor{cyan!10!white}54.65  \\
OLMo-3-7B-SFT & \cellcolor{cyan!10!white}63.95\,{\scriptsize(-32.10)} & \cellcolor{cyan!10!white}65.32 & \cellcolor{cyan!10!white}64.48 & \cellcolor{cyan!10!white}51.61 & \cellcolor{cyan!10!white}72.07 & \cellcolor{cyan!10!white}68.37 & \cellcolor{cyan!18!white}57.85 \\
\bottomrule
\end{tabular}
\end{table*}

\begin{figure}
    \centering
    \includegraphics[width=0.95\linewidth]{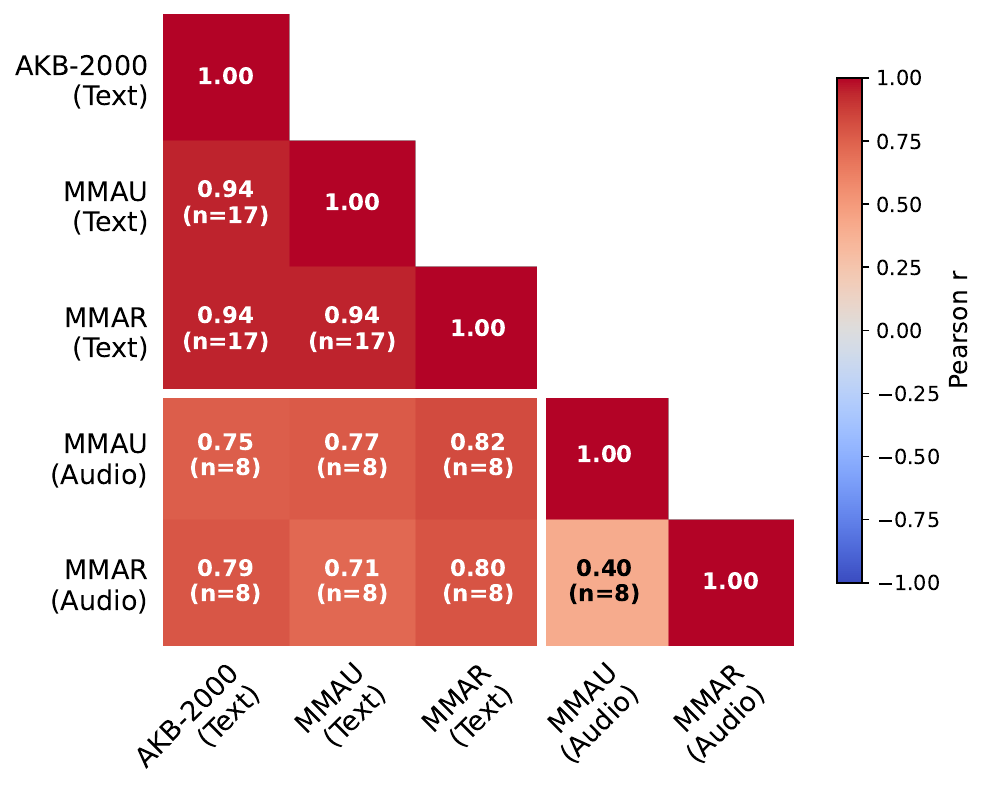}
    \caption{Pearson correlation heatmap across all five evaluation metrics.  The white line separates text-only metrics (top-left) from audio-grounded metrics (bottom-right).}
    \label{fig:correlation_heatmap}
\end{figure}

\subsection{Evaluation and Inference Setup}

Our AKB-2000 contains 2,000 four-option multiple-choice questions across 6 categories and 48 subcategories. The correct answer is uniformly distributed across the four options, ensuring that random-chance performance remains at 25\% and that no positional bias inflates model scores. For MMAU, we use the test-mini subset of 1,000 questions, and MMAR contains 1,000 questions in total. We report results on the speech, sound, and music categories from each benchmark. All questions are presented in a zero-shot setting without in-context examples.

For text-only evaluation (Sections~\ref{sec:aqa} and~\ref{sec:captioner}), we serve all open-weight LLMs using vLLM~\cite{kwon2023efficient} with each model's default generation configuration. Proprietary models are evaluated through their respective provider APIs. We use multimodal-capable Gemini-2.5-Pro as the default captioner in cascade evaluation; additional results using Omni-Captioner~\cite{ma2025omni} and Whisper~\cite{radford2023robust} as alternative captioners are reported as reference.
For audio-grounded evaluation (Section~\ref{sec:audio-finetuning}), we evaluate the resulting LALMs on the same MMAU and MMAR benchmarks with audio inputs, following the same evaluation setup from~\cite{lu2025desta25}. Since model outputs may not conform to a fixed answer format, we employ GPT-4o as a judge to verify whether each response matches the ground-truth answer. All benchmarks use accuracy (\%) as the evaluation metric.

\section{Results}


\begin{table}[]
    \centering
    \caption{The five most challenging subcategories in AKB-2000, averaged over 12 open-weight LLMs (\%).}
    \begin{tabular}{l|c}
    \toprule
    \textbf{Subcategory} & \textbf{Avg.} \\
    \midrule
     Phonetics \& Phonology & 60.2 \\
     Stress \& Emphasis & 58.3 \\
     Music Theory & 52.9 \\
     Syllable \& Stress & 49.1 \\
     Rhyme & 48.7 \\

    \bottomrule
    \end{tabular}
    
    \label{tab:hard_subcategories}
\end{table}

\subsection{Overall Trend}
\label{sec:overall_trend}

Figure~\ref{fig:correlation_heatmap} presents the Pearson correlation heatmap across all five evaluation metrics to provide a holistic view before examining individual results. Within the text-only block, all pairwise correlations reach 0.94, indicating that model rankings remain highly consistent across AKB-2000 and cascade evaluation, suggesting auditory knowledge is a coherent property of LLMs rather than an artifact of any single benchmark. Correlations between text-only and audio-grounded metrics are also strong (from $r=0.71$ to $r=0.82$). The one exception is the correlation between audio-grounded MMAU and MMAR ($r = 0.40$), which likely reflects limitations in our training data coverage, a point we will further discuss in Section~\ref{sec:audio_finetuning_results}.

Tables~\ref{tab:aqa_results}, \ref{tab:text_results}, and \ref{tab:audio_results_small} present the detailed results. Across all benchmarks, clear and consistent performance gaps emerge. Proprietary models form the top tier, with Gemini-2.5-Pro, GPT-5, and Claude-Sonnet-4.5 all exceeding 94\% on AKB-2000 and 69\% on both MMAU(Text) and MMAR(Text). Among open-weight models, AKB-2000 scores range from 45.90\% (Llama-2-7B) to 86.35\% (Phi-4-14B), and even at comparable 7–8B scale, Qwen2.5-7B (80.70\%) leads Llama-3.1-8B (68.10\%) by a wide margin despite similar release dates. The similar ordering persists across both text-only and audio-grounded evaluation, suggesting that text-only performance is largely predictive of how models rank after fine-tuning on real audio.

Among open-weight models, the Qwen family occupies the top tier across all three evaluations, with Qwen3-14B and Qwen2.5-7B achieving 85.05\% and 80.70\% on AKB-2000 and maintaining this advantage after audio fine-tuning. Interestingly, performance within Qwen3 shows no clear scaling trend with parameter count, as Qwen3-4B and Qwen3-8B perform at a comparable level in both text-only and audio-grounded settings. Phi-4-14B slightly outperforms Qwen3-14B at the top end, but the Phi family exhibits a steep drop at smaller scales, with Phi-4-mini falling significantly behind.
In contrast, the Llama and OLMo families occupy the lower tier across all settings. Within Llama, the newer Llama-3.1-8B underperforms its predecessor Llama-3-8B, indicating that model updates do not necessarily bring stronger auditory knowledge. OLMo reveals a different pattern, where post-training alignment stages (SFT $\rightarrow$ DPO $\rightarrow$ Instruct) yield noticeable gains on direct knowledge evaluation, yet these gains largely vanish in cascade evaluation, where all three checkpoints converge to a comparable level.%

\begin{table*}[t]
\centering
\caption{
Text-only cascade evaluation on MMAU and MMAR benchmark (\%). $\dagger$~denotes official cascade baselines from each benchmark~\cite{sakshi2025mmau,ma2025mmar}.\colorbox{cyan!20!white}{Shading} indicates relative ranking among open-weight models; values in parentheses show the gap from Gemini-2.5-Pro.}
\label{tab:text_results}
\begin{tabular}{l|c|ccc|c|ccc}
\toprule
\multirow{2}{*}{\textbf{Model}} & \multicolumn{4}{c}{\textbf{MMAU (Text)}}  & \multicolumn{4}{|c}{\textbf{MMAR (Text)}} \\
\cmidrule{2-5} \cmidrule{6-9}
& \textbf{Avg.} & Sound & Music &  Speech &  \textbf{Avg.} & Sound & Music & Speech \\
\midrule
\multicolumn{9}{l}{\textit{Captioner Comparison (LLM: Gemini-2.5-Pro)
}} \\
Official Baselines$^\dagger$ & 57.3 & 57.35 & 49.70 & 64.86 &  50.7 & 46.1& 40.3& 60.9 \\
Whisper-large-v3 & 61.7 & 52.55 & 53.59 & 78.98 & 61.6 & 41.21 & 44.66 & 77.55 \\
Omni-captioner & 68.9 & 69.97 & 61.08 & 75.68 & 65.5 & 51.52 & 46.12 & 77.21 \\
Gemini-caption & 70.9 & 68.77 & 66.47 & 77.48 & 71.8 & 64.85 & 49.03 & 81.97\\

\midrule
\midrule
\multicolumn{9}{l}{\textit{Gemini-caption + Proprietary LLM}} \\
Gemini-2.5-Pro & 70.9 & 68.77 & 66.47 & 77.48 & 71.8 & 64.85 & 49.03 & 81.97\\
GPT-5 & 71.9 & 72.97 & 66.47 & 76.28 & 69.8 & 66.06 & 47.09 & 80.27 \\
Claude-Sonnet-4.5 & 70.8 & 68.47 & 63.77 & 80.18 & 70.5 & 60.61 & 50.49 & 81.29 \\
Gemini-2.0-Flash & 69.6 & 68.47 & 63.77 & 76.58 & 64.4 & 58.18 & 47.09 & 75.85 \\
GPT-4o & 69.3 & 66.37 & 64.07 & 77.48 & 66.0 & 61.21 & 49.03 & 73.47 \\
\midrule
\multicolumn{9}{l}{\textit{Gemini-caption + Open-weight LLM}} \\
Qwen3-14B & \cellcolor{cyan!42!white}66.2\,{\scriptsize(-4.7)} & \cellcolor{cyan!46!white}65.77 & \cellcolor{cyan!34!white}61.08 & \cellcolor{cyan!46!white}71.77 & \cellcolor{cyan!50!white}64.3\,{\scriptsize(-7.5)} & \cellcolor{cyan!50!white}58.18 & \cellcolor{cyan!46!white}50.97 & \cellcolor{cyan!42!white}70.75 \\
Qwen3-8B & \cellcolor{cyan!50!white}66.8\,{\scriptsize(-4.1)} & \cellcolor{cyan!50!white}66.07 & \cellcolor{cyan!50!white}63.17 & \cellcolor{cyan!42!white}71.17 & \cellcolor{cyan!42!white}62.0\,{\scriptsize(-9.8)} & \cellcolor{cyan!50!white}58.18 & \cellcolor{cyan!26!white}41.26 & \cellcolor{cyan!50!white}72.79 \\
Qwen3-4B & \cellcolor{cyan!46!white}66.3\,{\scriptsize(-4.6)} & \cellcolor{cyan!42!white}62.46 & \cellcolor{cyan!46!white}62.28 & \cellcolor{cyan!50!white}74.17 & \cellcolor{cyan!34!white}61.0\,{\scriptsize(-10.8)} & \cellcolor{cyan!38!white}55.15 & \cellcolor{cyan!38!white}45.15 & \cellcolor{cyan!34!white}69.05 \\
Qwen2.5-7B & \cellcolor{cyan!38!white}64.5\,{\scriptsize(-6.4)} & \cellcolor{cyan!30!white}60.96 & \cellcolor{cyan!46!white}62.28 & \cellcolor{cyan!38!white}70.27 & \cellcolor{cyan!38!white}61.4\,{\scriptsize(-10.4)} & \cellcolor{cyan!38!white}55.15 & \cellcolor{cyan!42!white}47.57 & \cellcolor{cyan!46!white}71.77 \\
\midrule
Llama-3.1-8B & \cellcolor{cyan!14!white}53.6\,{\scriptsize(-17.3)} & \cellcolor{cyan!26!white}60.36 & \cellcolor{cyan!10!white}50.30 & \cellcolor{cyan!10!white}50.15 & \cellcolor{cyan!14!white}51.6\,{\scriptsize(-20.2)} & \cellcolor{cyan!18!white}50.91 & \cellcolor{cyan!18!white}37.86 & \cellcolor{cyan!14!white}57.48 \\
Llama-3-8B & \cellcolor{cyan!10!white}53.5\,{\scriptsize(-17.4)} & \cellcolor{cyan!10!white}51.65 & \cellcolor{cyan!14!white}52.40 & \cellcolor{cyan!18!white}56.46 & \cellcolor{cyan!30!white}54.4\,{\scriptsize(-17.4)} & \cellcolor{cyan!30!white}53.33 & \cellcolor{cyan!34!white}42.72 & \cellcolor{cyan!22!white}59.18 \\
Llama-2-7B & \cellcolor{cyan!5!white}43.2\,{\scriptsize(-27.7)} & \cellcolor{cyan!5!white}46.55 & \cellcolor{cyan!5!white}46.71 & \cellcolor{cyan!5!white}36.34 & \cellcolor{cyan!5!white}47.1\,{\scriptsize(-24.7)} & \cellcolor{cyan!14!white}50.30 & \cellcolor{cyan!5!white}33.01 & \cellcolor{cyan!5!white}50.68 \\
\midrule
Phi-4-14B & \cellcolor{cyan!34!white}62.9\,{\scriptsize(-8.0)} & \cellcolor{cyan!42!white}62.46 & \cellcolor{cyan!38!white}61.98 & \cellcolor{cyan!34!white}64.26 & \cellcolor{cyan!46!white}62.6\,{\scriptsize(-9.2)} & \cellcolor{cyan!50!white}58.18 & \cellcolor{cyan!50!white}51.46 & \cellcolor{cyan!38!white}69.73 \\
Phi-4-mini-4B & \cellcolor{cyan!18!white}56.1\,{\scriptsize(-14.8)} & \cellcolor{cyan!14!white}53.45 & \cellcolor{cyan!30!white}57.49 & \cellcolor{cyan!26!white}57.36 & \cellcolor{cyan!30!white}54.4\,{\scriptsize(-17.4)} & \cellcolor{cyan!26!white}52.73 & \cellcolor{cyan!30!white}41.75 & \cellcolor{cyan!30!white}61.22 \\
\midrule
OLMo-3-7B & \cellcolor{cyan!26!white}57.7\,{\scriptsize(-13.2)} & \cellcolor{cyan!34!white}61.26 & \cellcolor{cyan!18!white}55.69 & \cellcolor{cyan!14!white}56.16 & \cellcolor{cyan!22!white}53.2\,{\scriptsize(-18.6)} & \cellcolor{cyan!22!white}52.12 & \cellcolor{cyan!14!white}33.50 & \cellcolor{cyan!26!white}60.20 \\
OLMo-3-7B-DPO & \cellcolor{cyan!30!white}58.0\,{\scriptsize(-12.9)} & \cellcolor{cyan!18!white}58.86 & \cellcolor{cyan!26!white}56.29 & \cellcolor{cyan!30!white}58.86 & \cellcolor{cyan!18!white}52.4\,{\scriptsize(-19.4)} & \cellcolor{cyan!14!white}50.30 & \cellcolor{cyan!14!white}33.50 & \cellcolor{cyan!22!white}59.18 \\
OLMo-3-7B-SFT & \cellcolor{cyan!22!white}57.3\,{\scriptsize(-13.6)} & \cellcolor{cyan!22!white}59.16 & \cellcolor{cyan!22!white}55.99 & \cellcolor{cyan!22!white}56.76 & \cellcolor{cyan!14!white}51.6\,{\scriptsize(-20.2)} & \cellcolor{cyan!5!white}46.67 & \cellcolor{cyan!22!white}39.32 & \cellcolor{cyan!10!white}54.42 \\
\bottomrule
\end{tabular}
\end{table*}

\subsection{Results on Auditory Knowledge Benchmark Evaluation}

Table~\ref{tab:aqa_results} presents AKB-2000 performance across 6 categories. Since several proprietary LLMs were involved in the data curation process, their near-saturated performance (94.35--96.05\%) should be interpreted as an approximate upper bound rather than an unbiased comparison against other models. Accordingly, we treat Gemini-2.5-Pro as a reference point when discussing relative gaps in the table. Among proprietary models, earlier versions such as GPT-4o and Gemini-2.0-Flash score slightly below the most recent models. Across open-weight models, per-category rankings remain largely consistent regardless of model scale: a model that scores higher overall consistently shows superior performance across every individual category. This suggests that auditory knowledge in text-only LLMs reflects general language modeling capability rather than category-specific training data, since no model family shows an advantage in any particular domain within AKB-2000. A notable exception is Phonetic accuracy, which lags behind all other categories by 10--15 percentage points across every model family, revealing a systematic deficit that persists even among the strongest open-weight models.

At a finer granularity, Table~\ref{tab:hard_subcategories} lists the five most challenging subcategories, averaged over 12 open-weight models. Interestingly, most models struggle with tasks requiring knowledge of how words and sentences sound when spoken aloud. Four of the five most challenging subcategories require reasoning about pronunciation, prosody, or phonological structure that is not directly observable in written text. 
For the phonological subcategories, most models failed on questions such as ``\textit{Do the words `cat' and `hat' form a perfect rhyme?}'' or ``\textit{Which pair of words are homophones?}''. This failure pattern reveals a fundamental limitation of text-only pre-training: while LLMs learn rich semantic correlations between tokens, they are never exposed to the acoustic realization of language and therefore cannot ground their representations in how words actually sound. Humans, who routinely experience spoken words in daily life, can naturally recognize that ``\textit{flour}'' and ``\textit{flower}'' sound alike despite being semantically distinct. The primary objective of LLM training, however, is to learn semantic correlations between tokens, which is sufficient for retrieving factual and linguistic knowledge but fails to capture phonological associations. 

Consequently, LLMs systematically fail to associate words that sound alike but look different, and this gap may directly undermine performance in downstream speech applications such as spoken dialogue systems and ASR, where phonological competence is essential. Addressing this limitation will likely require phonology-aware training strategies, such as incorporating pronunciation lexicons or phoneme-level supervision, that go beyond what standard text corpora can provide.

\begin{table}[t]
\centering
\caption{Audio-grounded MMAU and MMAR performance (\%). The upper block lists published LALM systems; the lower block reports our fine-tuned LALMs with DeSTA framework.}
\label{tab:audio_results_small}
\begin{tabular}{l|c|c}
\toprule
\multirow{2}{*}{\textbf{Model}} & \textbf{MMAU} & \textbf{MMAR} \\
& \textbf{(Audio)} & \textbf{(Audio)}\\
\midrule
\multicolumn{3}{l}{\textit{Published Systems}} \\
Gemini-2.5-Pro (Audio) & 71.60 & 74.70 \\
Audio Flamingo 3~\cite{goel2025audio}      & 73.30 & 58.60 \\
Qwen2.5-Omni~\cite{xu2025qwen3}         & 71.50 & 56.70 \\
DeSTA2.5-Audio~\cite{lu2025desta25}       & 66.00 & 50.80 \\
Phi-4-mm~\cite{abouelenin2025phi}         & 65.70 & 40.20 \\
\midrule
\multicolumn{3}{l}{\textit{Fine-tuned LALM (ours)}} \\
Qwen3-14B  & \cellcolor{cyan!46!white}66.20 & \cellcolor{cyan!46!white}52.90 \\
Phi-4-14B   & \cellcolor{cyan!26!white}61.10 & \cellcolor{cyan!42!white}52.50 \\
\midrule
Qwen3-8B     & \cellcolor{cyan!30!white}61.70 & \cellcolor{cyan!50!white}53.00 \\
Qwen2.5-7B   & \cellcolor{cyan!50!white}66.60 & \cellcolor{cyan!22!white}47.30 \\
Llama-3.1-8B  & \cellcolor{cyan!5!white}56.40 & \cellcolor{cyan!22!white}47.70 \\
OLMo-3-7B     & \cellcolor{cyan!10!white}56.90 & \cellcolor{cyan!10!white}44.90 \\
\midrule
Qwen3-4B    & \cellcolor{cyan!34!white}62.90 & \cellcolor{cyan!30!white}49.20 \\
Phi-4-mini-4B & \cellcolor{cyan!26!white}61.00 & \cellcolor{cyan!5!white}44.20 \\
\bottomrule
\end{tabular}
\end{table}

\subsection{Results on Cascade Evaluation}

Table~\ref{tab:text_results} presents cascade evaluation results on the MMAU and MMAR benchmarks, where audio is first converted to text descriptions by a captioner before being passed to a text-only LLM. The relative ranking of models follows a similar trend as in the AKB-2000 evaluation, suggesting that an LLM's inherent auditory knowledge consistently influences its downstream reasoning ability regardless of the evaluation format.

In the cascade paradigm, the captioner plays a critical role, as recognition errors in the caption propagate directly to the downstream LLM. While this error propagation can be problematic when building robust agentic systems, our goal here is to isolate the reasoning capability of different LLMs by providing them with identical caption inputs and measuring the relative strength between LLMs. We additionally include several captioners alongside the official leaderboard baselines in Table~\ref{tab:text_results}.
Among them, using Gemini-2.5-Pro as the captioner surpasses all other configurations by a large margin, scoring 70.90\% and 71.80\% on MMAU and MMAR respectively. 
This strong captioner performance ensures that the text descriptions retain sufficient auditory detail for downstream reasoning. At the same time, the cascade results reveal that captioner quality remains a critical bottleneck. Even when pairing the strongest captioner (Gemini-2.5-Pro) with a top-tier proprietary LLM, overall performance plateaus at around 70\% on both benchmarks, leaving substantial room for improvement. 

Breaking down by category, most Qwen models show a disproportionate advantage in Speech, exceeding 70\% on both MMAU and MMAR while other open-weight families of comparable scale remain around 50–60\%. This suggests that Qwen may encode more speech-related knowledge through its text-only pre-training, such as understanding of speaker attributes, prosody, and conversational structure. In contrast, the Sound and Music categories show considerably smaller differences across open-weight families.
Notably, Qwen3-14B (66.20\% on MMAU, 64.30\% on MMAR) approaches or matches Gemini-2.0-Flash (69.60\% and 64.40\%), suggesting that capable open-weight LLMs can already close the gap with earlier proprietary models under the cascade setting.


\begin{figure}[t]
     \centering
     \begin{subfigure}[b]{0.95\linewidth}
         \centering
         \includegraphics[width=0.87\linewidth]{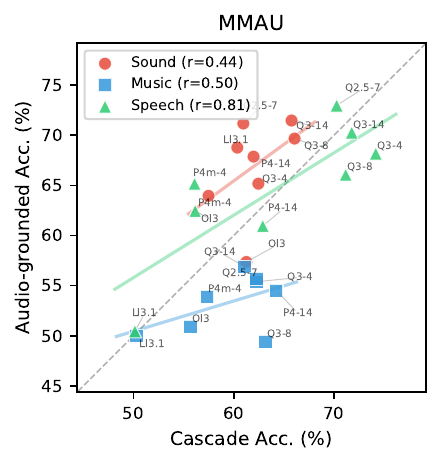}
         \label{fig:left}
     \end{subfigure}
     \begin{subfigure}[b]{0.95\linewidth}
         \centering
         \includegraphics[width=0.87\linewidth]{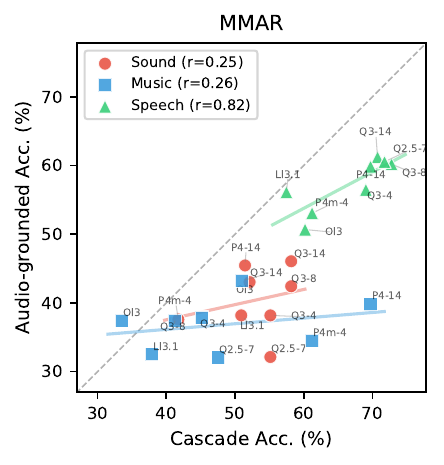}
         \label{fig:right}
     \end{subfigure}
     
     \caption{Category-level scatter plots comparing cascade and audio-grounded accuracy (\%) for 8 fine-tuned LALMs, broken down by Sound, Music, and Speech domains.}
     \label{fig:mmau_mmar_category}
\end{figure}

\subsection{Results on Audio-grounded Evaluation}
\label{sec:audio_finetuning_results}

Based on the text-only evaluation results, we fine-tune 8 open-weight LLMs across different parameter scales (4B--14B) and model families into LALMs. The selection is guided by the tier structure observed in text-only evaluation: we include top-tier models (Qwen3-14B, Phi-4-14B), mid-tier models (Qwen3-8B, Qwen3-4B, Qwen2.5-7B), and lower-tier models (Llama-3.1-8B, OLMo-3-7B, Phi-4-mini-4B), ensuring that the fine-tuning experiment covers the full performance spectrum.

As presented in Table~\ref{tab:audio_results_small}, the best and worst fine-tuned models differ by over 10 points on MMAU and 8 points on MMAR, confirming that the choice of backbone LLM remains a significant factor even when all other components are held constant. Notably, Qwen2.5-7B and Qwen3-14B achieve the highest MMAU scores (66.6\% and 66.2\%, respectively), matching or surpassing DeSTA2.5-Audio (66.0\%), which uses Llama-3.1-8B as its backbone with ten times the training data~\cite{lu2025desta25}. This suggests that selecting a stronger backbone can compensate for a large gap in training data scale.

These findings also expose a broader challenge in comparing LALMs across the literature. Systems reported in prior work differ simultaneously along multiple axes, including backbone LLM, training data, audio encoder, and training recipe, making it difficult to attribute performance gains to any single factor. In particular, a stronger backbone alone can account for a substantial portion of the observed improvement, a confound that is rarely acknowledged. Our controlled ablation, which isolates the backbone LLM as the sole variable, establishes backbone selection as a first-order design decision that warrants explicit consideration in future LALM development.

As shown in Tables~\ref{tab:text_results} and ~\ref{tab:audio_results_small}, the performance gap between cascade and audio-grounded results further reveals where current end-to-end architectures fall short. Cascade pipelines leveraging a strong captioner already match or surpass Audio Flamingo 3 and Qwen2.5-Omni on MMAR (e.g., 62.0\% for Qwen3-8B versus 58.6\% and 56.7\%). This finding points to a potential audio-text alignment bottleneck in end-to-end LALMs, where the audio encoder may fall short of preserving fine-grained details that a specialized captioner can explicitly articulate, resulting in information loss at the multimodal training stage.

We observe a similar bottleneck in our controlled experiments. Figure~\ref{fig:mmau_mmar_category} illustrates the performance within categories on MMAU and MMAR under cascade and audio-grounded evaluation, revealing that the text-to-audio transfer pattern differs substantially across domains. 
Among the three domains, speech is by far the most represented in our training data, while sound and music account for a smaller proportion. 
For speech, the correlation between cascade and audio-grounded performance remains strong ($r=0.81$), and the performance spread across models is largely preserved in both benchmarks. In contrast, the sound and music domains exhibit weaker correlations that flatten meaningful differences between models, especially on MMAR, suggesting that the audio module, rather than the backbone LLM, becomes the primary bottleneck when training data is insufficient. This category-level disparity also accounts for the low cross-benchmark correlation ($r=0.40$) between audio-grounded MMAU and MMAR observed in Section~\ref{sec:overall_trend}, as the two benchmarks differ in their relative difficulty, making model rankings sensitive to training data coverage in those domains. 
Therefore, designing more targeted data sources or training recipes will be necessary to fully leverage the inherent capability of the language model backbone across all audio domains.

\section{Conclusion}

In this work, we present a holistic evaluation of text-only LLMs and reveal how auditory knowledge in LLM backbones shapes LALMs. We evaluate the models under text-only and multimodal settings: direct knowledge probing on a curated auditory knowledge benchmark (AKB-2000), cascade evaluation, and audio-grounded evaluation via end-to-end fine-tuning. Our results show that auditory knowledge varies substantially across model families, and text-only performance is strongly correlated with audio-grounded performance, making AKB-2000 and cascade evaluation a lightweight proxy for LLM selection in LALM research. These findings establish the choice of LLM backbone as a first-order design decision that warrants explicit consideration in future LALM development. We further identify phonological reasoning as a systematic blind spot of text-only pre-training, and show that the cascade pipeline performs comparably to, or even surpasses, recent state-of-the-art LALMs, suggesting that most LALMs do not fully utilize the inherent capabilities of the LLM backbone. Taken together, our findings demonstrate that the auditory knowledge encoded in the LLM backbone fundamentally shapes every stage of an audio understanding system, and that a holistic understanding of this knowledge is essential for building stronger LALMs.

\section{Generative AI Use Disclosure}
In this work, generative AI tools were used in two capacities. First, generative AI tools were used to assist in proofreading and improving the fluency of the manuscript. Second, as described in Section~\ref{sec:aqa}, LLMs were employed to assist in curating candidate questions for AKB-2000, following detailed human-authored guidelines, with all questions verified by human annotators before inclusion. All scientific content, experimental design, analysis, and conclusions are solely the work of the authors.





\section{Acknowledgments}
We acknowledge the computational and storage support provided by the National Center for High-performance Computing (NCHC) of the National Applied Research Laboratories (NARLabs) in Taiwan.

\bibliographystyle{IEEEtran}
\bibliography{mybib}

\end{document}